\documentclass[conference]{IEEEtran}
\IEEEoverridecommandlockouts
\usepackage{cite}
\usepackage{amsmath,amssymb,amsfonts}
\usepackage{algorithmic}
\usepackage{graphicx}
\usepackage{textcomp}
\usepackage{xcolor}
\def\BibTeX{{\rm B\kern-.05em{\sc i\kern-.025em b}\kern-.08em
    T\kern-.1667em\lower.7ex\hbox{E}\kern-.125emX}}
    
\def \akis [#1]{\textcolor{red}{AP: #1}}  
\def \sharath [#1]{\textcolor{blue}{SA: #1}}  

\def\xp {\tilde{\mathbf{x}}}
\def\Xp {\tilde{\mathbf{X}}}

\def\xr {\mathbf{x}}
\def\Xr {\mathbf{X}}
\def\D {\mathbf{D}}
\def\A {\mathbf{A}}
\def\tA {\tilde{\mathbf{A}}}

\begin{document}

\title{Differentiable Tracking-Based Training of Deep Learning Sound Source Localizers
}

\author{\IEEEauthorblockN{Sharath Adavanne\textsuperscript{*}, Archontis Politis\textsuperscript{*}, Tuomas Virtanen \thanks{\textsuperscript{*} Equally contributing authors in this paper.}}
\IEEEauthorblockA{\textit{Audio Research Group, Tampere University}\\
Tampere, Finland \\
{name.surname}@tuni.fi}
}

\maketitle

\begin{abstract}
Data-based and learning-based sound source localization (SSL) has shown promising results in challenging conditions, and is commonly set as a classification or a regression problem. Regression-based approaches have certain advantages over classification-based, such as continuous direction-of-arrival estimation of static and moving sources. However, multi-source scenarios require multiple regressors without a clear training strategy up-to-date, that does not rely on auxiliary information such as simultaneous sound classification. We investigate end-to-end training of such methods with a technique recently proposed for video object detectors, adapted to the SSL setting. A differentiable network is constructed that can be plugged to the output of the localizer to solve the optimal assignment between predictions and references, optimizing directly the popular CLEAR-MOT tracking metrics. Results indicate large improvements over directly optimizing mean squared errors, in terms of localization error, detection metrics, and tracking capabilities.
\end{abstract}

\begin{IEEEkeywords}
sound source localization, deep-learning acoustic processing, multi-target tracking \end{IEEEkeywords}

\section{Introduction}

    Sound source localization (SSL) has been one of the most classic and consistently researched topics of microphone array signal processing \cite{brandstein2001microphone}, with wide ranging applications from acoustic scene analysis \cite{politis2020overview} and acoustic monitoring \cite{valenzise2007scream}, to speech enhancement \cite{dibiase2001robust} and spatial audio rendering \cite{pulkki2018parametric}. SSL methods usually focus on providing the direction-of-arrival (DOA) of a single or multiple concurrent sources, while temporal smoothing of a single DOA and association  of multiple estimates of multiple DOAs over time forms the topic of sound source tracking (SST) \cite{dibiase2001robust}. Recently, the field, traditionally dominated by geometric or statistical model-based approaches, has seen a surge in data- and learning-based SSL proposals using deep neural network (DNN) architectures \cite{wang2018robust, adavanne2018direction, adavanne2018sound, perotin2019crnn, chakrabarty2019multi, nguyen2020robust, diaz2020robust, bianco2020semi}.
    
    A deep-learning paradigm on SSL opens up a few interesting research questions, such as basic spectrogram\cite{adavanne2018sound, chakrabarty2019multi} versus refined spatial \cite{perotin2019crnn, nguyen2020robust}  multichannel input features, coupling the network architecture to SSL effectively \cite{chakrabarty2019multi, krause2021comparison}, choosing appropriate training source signals for generalization \cite{chakrabarty2019multi, vargas2021improved}, strong versus weak supervision \cite{bianco2020semi}, and posing SSL as a classification \cite{adavanne2018direction, chakrabarty2019multi, perotin2019crnn, nguyen2020robust} or regression \cite{adavanne2018sound, diaz2020robust, perotin2019regression} problem. The latter division was already present in earlier attempts of single-source deep-learning SSL, such as classification in \cite{xiao2015learning} and regression in \cite{vesperini2016neural}. In classification-based SSL, the range of possible DOAs is discretized into distinct DOA classes, with the classifier having as many outputs as the number of them. Classification-based SSL has certain advantages: it can serve as a simultaneous source activity detector and it can handle multiple sources with a single network architecture. On the other hand, the gridding determines the effective resolution, errors are higher at boundaries between grid points, and coarse resolutions cannot accommodate well moving source scenarios. Additionally, for full 3D DOA estimation in azimuth-elevation, even moderate resolutions require hundreds of classes, posing challenges in obtaining adequate training data and training effectively.
    
    Classification-based SSL was the dominant paradigm until recently, where studies such as \cite{adavanne2018sound} brought increased attention to regression, with similar performance to classification further validated, e.g., in \cite{perotin2019regression}. Regression-based SSL has its own advantages: a single regressor on DOA vectors or angles can handle the whole DOA domain for a single source with one to three outputs, estimation is continuous, and moving source scenarios are handled naturally \cite{adavanne2019localization, adavanneThesis}. However, some auxiliary activity detection is required to gate the constant stream of DOAs during inference \cite{diaz2020robust}. Furthermore, in the multi-source case, as many regressors as the presumed maximum number of sources are needed, posing problems of permutations between sources and regression outputs, preventing effective training and increasing localization errors during inference \cite{Cao2020}.
    
    Regression-based SSL is popular in the context of joint sound event localization and detection (SELD), e.g., in the submissions of the DCASE 2019 and DCASE 2020 challenges \cite{politis2020overview}, where participants could use simultaneous event classification information to infer activity and disentangle permutation issues. 
    However, in a classical multi-source SSL setting independent of source signal type, not much work has been done in addressing the above issues.
    In this study, we propose a training strategy for multi-source regression-based SSL that circumvents all the aforementioned issues. More specifically, a) instead of optimizing only spatial localization errors as it is commonly done, source detection terms are included in the loss improving overall performance, b) permutation errors are avoided by integrating tracking-inspired loss terms, c) the method provides an end-to-end training strategy that can handle dynamic changing conditions with variable number of sources, suitable for real-life annotated recordings.

\section{Localization and tracking metrics}

Considering a recording with maximum number $N_\mathrm{max}$ sound sources active over its duration, not necessarily simultaneously, we can define the predictions of an SSL system as $\Xp_t = [\xp_1(t), ...,\xp_i(t), ..., \xp_{M_t}(t)]$, where $\xp = [\tilde{x},\tilde{y},\tilde{z}]$ is the estimated DOA or position vector of a single source, and $M_t$ is the number of predictions at the $t$-th frame. At the same time, $N_t\leq N_{max}$ ground truth sources and their locations are denoted by  $\Xr_t = [\xr_1(t), ...,\xr_j(t), ..., \xr_{N_t}(t)]$. The combinations of estimations and predictions form the $M_t\times N_t$ distance matrix $\D_t$ with an appropriate spatial distance measure for the application; e.g. the angular distance $d_{ij} = \arccos( \xp_i\cdot \xr_j / ||\xp_i|| ||\xr_j|| )$ when DOAs are considered. Based on $\D$, we can also consider an optimal association of references and predictions, in a minimum cost sense, expressed by a $M_t \times N_t$ binary association matrix $\A_t = \mathcal{H}(\D)$, where $\mathcal{H}(\cdot)$ is the Hungarian algorithm \cite{Hungarian}. The association matrix $\A$ allows an optimal frame-wise \emph{localization error} (LE) to be computed between the $K_t=\min(M_t,N_t)$ associated predictions-references, as
\begin{equation}
    LE_t = \frac{1}{K_t}\sum_{i,j} a_{ij}(t) d_{ij}(t) = \frac{|| \A_t \odot \D_t||_1}{||\A_t||_1},
    \label{eq:le}
\end{equation}
with $d_{ij} = [\D]_{ij}$, $a_{ij} = [\A]_{ij}$, $||\cdot||_1$ being the $L_{1,1}$ entrywise matrix norm, and $\odot$ the entrywise matrix product. Complementary to LE, the association matrix $\A$ indicates hits/true positives (TP) $TP_t = K_t$, false alarms/false positives (FP) $FP_t = \max(0, M_t-N_t)$ , and misses/false negatives (FN) $FN_t = \max(0, N_t-M_t)$. From those, detection metrics such as the \emph{localization recall} (LR), \emph{localization precision} (LP), and a \emph{localization F1-score} (LF1) can be computed \cite{politis2020overview}.

The above SSL metrics reveal the performance of the system in detecting and localizing accurately the sources in the scene but not how well the estimates are maintained across time, which is the task of tracking. Tracking metrics for multiple objects or sources is still an open field of research. Some established ones, such as OSPA \cite{schuhmacher2008consistent} favour trajectory consistency, while others like the CLEAR Multiple Object Tracking (MOT) metrics \cite{bernardin2008evaluating} try to balance between good localization performance in presence of \emph{identity switches} (IDS), and consistent identities between estimates from frame-to-frame. Two complementary MOT metrics are proposed in \cite{bernardin2008evaluating}, the MOT-precision (MOTp), and MOT-accuracy (MOTa)
\begin{align}
    MOTp &= \frac{\sum_t || \A_t \odot \D_t||_1}{\sum_t K_t} \\
    MOTa &= 1 - \frac{\sum_t FP_t+FN_t+IDS_t}{\sum_t N_t}.
\end{align}
As it is evident, MOTp is actually equivalent to LE, averaged across all frames. IDS can be computed by comparison of the current and previous frame association matrices $\A_t, \A_{t-1}$ and knowledge of the source ID for every column of $\A$ across frames, e.g. as in \cite{xu2020train}. MOTa itself is a combination of detection metrics with an additional tracking penalty expressed by IDS.

\begin{figure}[!tp]
\centerline{\includegraphics[width=\linewidth, height=8cm,keepaspectratio]{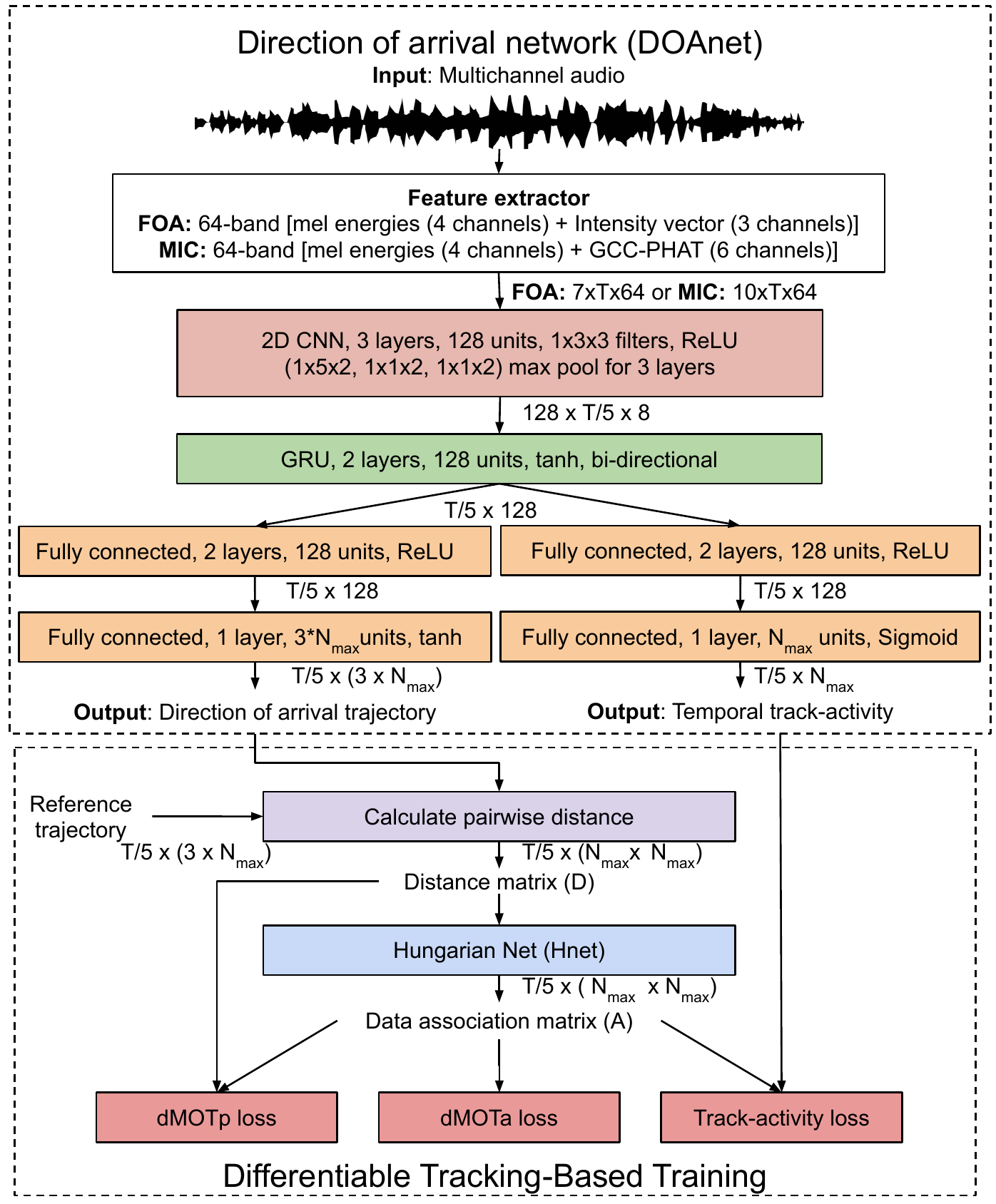}}
\vspace{-10pt}
\caption{Block diagram of Differentiable Tracking-Based Training.}
\label{fig:doanet}
\vspace{-10pt}
\end{figure}

\section{Proposed method}

The proposed method is strongly inspired by the work of \cite{xu2020train} on training video object detectors with an additional network plugged in the end of the object detectors, optimizing directly the MOT metrics through a differentiable soft-approximation of them. To the best of our knowledge, this strategy has not been attempted before on SSL problems, and its effects on multi-source regression have not been studied. Our proposal follows the training of \cite{xu2020train} with certain modifications. The overall block diagram is shown in Fig.~\ref{fig:doanet}, consisting of the localization network, termed herein \emph{DOAnet}, and a deep Hungarian network (\emph{Hnet}) taking as input the distance matrix $\mathbf{D}$ computed from the DOAnet outputs, and predicting an association matrix $\tilde{\mathbf{A}}$. The $\tilde{\cdot}$ indicates a (soft) differentiable approximation of the underlying quantity. A series of differentiable matrix manipulations follow that provide  further soft approximations of $\tilde{LE}, \tilde{FP}, \tilde{TP}, \tilde{FP}, \tilde{FN}$, and $\tilde{IDS}$. From those approximations, the differentiable $dMOTp$ and $dMOTa$ are constructed and their combination serves as the overall training objective. A difference with the video-based work of \cite{xu2020train} is that, contrary to video object detectors, the localization regressors are constantly active. Hence, we introduce an additional track activity output branch in the localizer, contributing a third loss term in the overall loss. During inference, the DOA and track activity outputs are combined to form consistent DOA trajectories.


\begin{figure}[!tp]
\centerline{\includegraphics[width=\linewidth, height=5cm,keepaspectratio]{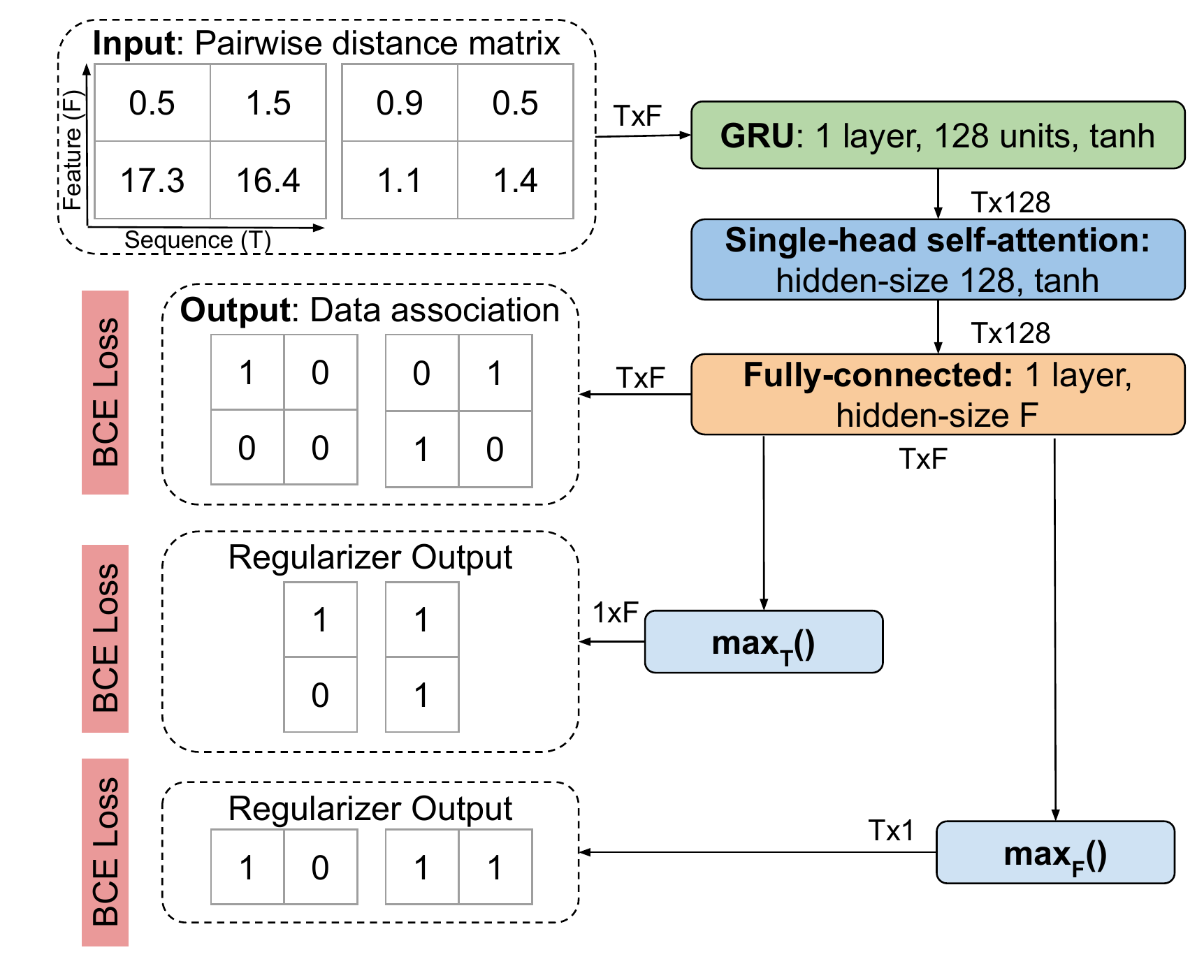}}
\vspace{-10pt}
\caption{Block diagram of Hungarian network.}
\label{fig:hnet}
\vspace{-10pt}
\end{figure}

\subsection{Hungarian network (Hnet)}
\label{sec:hnet}

The Hnet is the fundamental block of the proposed differentiable tracking-based training strategy. It estimates the association matrix $\tA$ of a dimension identical to the input distance matrix $\D$. In comparison to the deep Hungarian network proposed in~\cite{xu2020train}, we employ a simplified architecture as shown in Fig.~\ref{fig:hnet} with three losses to train Hnet swiftly and efficiently. We use a gated recurrent unit (GRU) input layer with 128 units, that treats one of the two dimensions of the input matrix D as the time-sequence, and the other as the feature length. The output time-sequence of GRU is fed to a single-head self-attention network~\cite{vaswani2017attention} to identify the time steps with correct associations. The output of the self-attention layer is processed by a fully-connected network with a sigmoid non-linearity, that estimates $\tA$ as a multiclass multilabel classification task. 

Additionally, to guide the network to predict a maximum of one association per row and column, as expected for associations resulting from the Hungarian algorithm; we perform max-operation on the output of fully-connected network (before the sigmoid non-linearity used to compute $\tA$) along both temporal ($\bf{max_T()}$) and feature ($\bf{max_F()}$) axes. We employ sigmoid non-linearity on these outputs, since more than one class can be active in an output instance. Finally, the Hnet is trained in a multi-task framework with weighted combinations of the three losses, each computed using binary cross-entropy between the predictions and the target labels of $\A$, $\bf{max_T(\A)}$, and $\bf{max_F(\A)}$ respectively.

\subsection{Differentiable direction of arrival network (DOAnet)}
\label{sec:doanet}

Regarding the DOAnet, we propose a convolutional recurrent neural network (CRNN) architecture, following an updated version of SELDnet~\cite{adavanne2018sound} as the baseline of DCASE 2020~\cite{politis2020dataset}. The detailed architecture is shown in Fig.~\ref{fig:doanet}. Based on the chosen array type, we employ different multichannel acoustic features. For the first-order Ambisonics (FOA) format we extract 4 channel-wise mel-band energies and 3 channels of acoustic active intensity vectors \cite{pulkki2018parametric} representing their $(x, y,z)$ vector components, resulting to in total 7 features. All features are computed using 64 mel-bands resulting in a total feature dimension of $7\times T\times 64$, where $T$ is the number of temporal input frames. Similarly, for the MIC array we compute 4 channel-wise mel energies, and GCC-PHAT curves between channel-pairs resulting in 6-channels of features, and a total feature dimension of $10\times T\times 64$.
        
The network is identical for both spatial formats. Three convolutional layers, with 128 units each, are employed to learn shift-invariant features from the input acoustic features. Maxpooling is performed on both temporal and feature axes to obtain an output of dimension $128\times T/5\times 8$, where $T/5$ amounts to 100 msec and is equal to the temporal resolution of DOA labels in the dataset (see Section~\ref{sec:dataset}). Two layers of bidirectional GRUs, each with 128 units are employed to model the temporal structure of the convolutional features. Thereafter, two separate branches are employed to learn - a) the  DOA trajectories and b) their temporal track activity. The DOA trajectory output branch is of dimensions $T/5 \times(3 N_{max})$, where for each time frame the location of $N_{max}$ DOAs in Cartesian form is estimated using regression. Since DOAs constitute unit vectors and their components are bounded in $[-1,1]$, tanh activations are used. The second output is of dimension $T/5\times N_{max}$, indicating track activity for the $N_{max}$ DOA outputs at each time instance. Since any of the $N_{max}$ tracks can be active for a given frame, sigmoid activations are used.

During training of the DOAnet, pairwise Euclidean distances are computed between the $M_t$ predicted and $N_t$ reference DOAs, forming the distance matrix $\D$. Euclidean distances are used instead of angular (cosine) distances, since they were found in \cite{adavanne2018sound, perotin2019regression} to perform better during training. Note that we embed the pairwise distances in a $\D$ matrix of the maximum dimensions $N_\mathrm{max} \times N_\mathrm{max}$, padding rows and columns beyond $M_t, N_t$ with out-of-range values (i.e. $>>2$). The input sequence to Hnet has finally the dimension $T/5\times N_{max}\times N_{max}$. A pre-trained Hnet with frozen weights is then employed to obtain the soft associations $\tA$ from input $\D$. The combined DOAnet, Hnet, and final differentiable operations forming dMOTa and dMOTp, are jointly trained by a weighted combination of three losses - the dMOTA, dMOTP, and the track-activity loss. Since the Hnet weights are frozen, weight updates are only performed on DOAnet.

The differentiable tracking losses of dMOTa and dMOTp are computed in an identical fashion as proposed in~\cite{xu2020train} using the inputs $\D$ and $\tA$. 
As the loss for the track-activity branch, we perform a row max operation on the $\tA$ matrix to obtain a $N_\mathrm{max} \times 1$ vector of soft activity values for all regressors. Higher values indicate higher probability of activity. The values are further thresholded and binarized. The collection of such vectors across frames result in the binary matrix $\mathbf{D}_\mathrm{ref}$ of size $T/5\times N_\mathrm{max}$ that is treated as the reference temporal activity of the DOA regressors. Then, the temporal activity branch is optimized with a binary cross entropy loss between its predicted $\mathbf{D}_\mathrm{pred}$ and reference $\mathbf{D}_\mathrm{ref}$ track activities.
In order to support open research and reproducibility we are publicly releasing the code of Hnet\footnote{https://github.com/sharathadavanne/hungarian-net} and DOAnet\footnote{https://github.com/sharathadavanne/doa-net}.

\begin{table}[t]
\large
\centering
\caption{Results of differentiable tracking based training on DCASE2020 SELD task dataset.}
\label{tab:results}
\resizebox{0.49\textwidth}{!}{%
\begin{tabular}{l|cccc|cccc}
& \multicolumn{4}{c|}{\textbf{FOA}} & \multicolumn{4}{c}{\textbf{MIC}} \\ 
\cline{2-9}
\textbf{Loss function} & \textbf{\begin{tabular}[c]{@{}c@{}}LE  $\downarrow$/\\ MOTp\end{tabular}} & \textbf{MOTa} $\uparrow$ &  \textbf{IDS} $\downarrow$ & \textbf{LR} $\uparrow$  & \textbf{\begin{tabular}[c]{@{}c@{}}LE  $\downarrow$/\\ MOTp\end{tabular}} & \textbf{MOTa} $\uparrow$&  \textbf{IDS} $\downarrow$ & \textbf{LR} $\uparrow$ \\
MSE     & 25.4 & $\sim$ & $\sim$ & $\sim$   & 25.3 & $\sim$ & $\sim$ & $\sim$ \\ 
dMOTp   & 13.7 & $\sim$ & $\sim$ & $\sim$   & 13.6 & $\sim$ & $\sim$ & $\sim$ \\
\multicolumn{9}{l}{\textbf{+Augmentation}} \\ \hline
dMOTp   & 12.1 & $\sim$ & $\sim$ & $\sim$   & 11.8 & $\sim$ & $\sim$ & $\sim$ \\
dMOTp+Act   & 9.7           & 69.0          & 2374  & 86.9          & 8.7          & 71.3 &     1982     & 87.3 \\
dMOTp+dMOTa+Act & 9.5 & \textbf{70.5} & \textbf{2188} & \textbf{88.1} & 8.5 & \textbf{72.1} & \textbf{1812} & \textbf{87.6} \\
\multicolumn{9}{l}{} \\
\multicolumn{9}{l}{\textbf{DCASE2020 top submissions}} \\ \hline
Du\_USTC (1)        & \textbf{7.4}    & $\sim$ & $\sim$  & 84.7  & \textbf{7.4}     & $\sim$ & $\sim$ & 84.7  \\
Nguyen\_NTU (2)     & 12.1 & $\sim$ & $\sim$  & 82.0  &  $\sim$   & $\sim$ & $\sim$ & $\sim$  \\
Shimada\_SONY (3)   & 7.5 & $\sim$ & $\sim$  & 83.5  &    $\sim$  & $\sim$ & $\sim$ & $\sim$ 
\end{tabular}
}
\vspace{-10pt}
\end{table}

\section{Evaluation}

\subsection{Hungarian network training}
\label{sec:training}

In order to train the Hnet, we generate a dataset with a training split of 405k distance matrices $\D$ and their corresponding association matrices $\A$. The validation split is 10\% the size of the training split. The dimensions of $\D$ and $\A$ are the same and fixed to ($N_{max} \times N_{max}$), where $N_{max} = 2$ is the maximum polyphony in the dataset.
We sample equal number of $\D$ matrices by randomly choosing reference and predicted DOAs from spherical equiangular grids with resolutions of 1, 2, 3, 4, 5, 10, 15, 20, and 30 degrees.
All combinations of (number of predictions, number of reference) such as {(0,0), (0,1), (1,0), (1,1), (1,2), (2,1), (2,2)} are represented equally in the dataset. As mentioned in Sec.~\ref{sec:doanet}, Euclidean distances are used to form the distance pairs in $\mathbf{D}$.

Due to padding $\D$ to $N_\mathrm{max}\times N_\mathrm{max}$ dimensions even when $M_t,N_t<N_\mathrm{max}$, random high distance values are assigned to the respective inactive entries, helping Hnet to easily identify the correct number of active DOAs and their associations. An example is depicted in the first input $\mathbf{D}$ distance matrix of Fig.~\ref{fig:hnet}, with the corresponding association $\A$ under it. After training, Hnet achieves an F-score of $>$99\% on any $\mathbf{D}$ data generated with the aforementioned specifications.

\subsection{Evaluation setup}
\label{sec:dataset}

For the evaluation of the whole differentiable training strategy we use the development set of the \emph{TAU-NIGENS Spatial Sound Events 2020} dataset \cite{politis2020dataset}, provided in the DCASE2020 Task 3 (SELD) challenge. It consists of diverse spatialized sound events, including moving sources, emulated in challenging real reverberant conditions using measured room impulse responses from 13 different rooms, with real spatial ambient noise added. The recordings are offered in two 4-channel formats: a tetrahedral microphone array (MIC), and first-order Ambisonics (FOA). The same development set split is used for training, validation, and testing as indicated in the challenge \cite{politis2020dataset}. The spatiotemporal annotations are used to extract the reference DOAs, event identities, and temporal activations at each frame, required for the evaluation of the system, ignoring the class/sound-type label of the original annotations.

An additional evaluation is conducted on an augmented version of the dataset. Following a simple spatial augmentation strategy popular in DCASE 2020 \cite{Du2020_task3_report}, additional recordings of overlapping sources were generated by simple mixing of recordings with no overlap with another four non-overlapping ones, resulting in 4 times the original dataset of 2-source overlapping recordings.

\section{Results}

The results across both formats, MIC and FOA are presented in Table~\ref{tab:results}. Results of $LE/MOTp$ are shown for all tested configurations, while results for $MOTa, IDS, LR$ are shown only for configurations including the track activity detection branch. Without activity detection, all regressors are constantly outputting DOAs, hence $LR=100\%$ and the rest of the detection scores are not meaningful. As the first result, and as a baseline, we train the DOAnet using an MSE loss between predicted and reference DoAs without any association strategy. This configuration ends up in large errors due to permutations on the estimates that prohibit effective training and result in suboptimal performance during inference. Just replacing it with the dMOTp loss, which finds the optimal assignment with the minimum frame-wise LE, almost doubles the localization accuracy. Moving to the augmented dataset for the same dMOTp loss, we have a further small decrease in LE. By introducing the activity detection branch and the respective loss, the LE/MOTp is further reduced below 10$^\circ$. With track activity information introduced, we can also get a realistic picture of the localization detection and MOTa scores. Solely the combination of track activity loss and dMOTp achieves a high $LR$ in the challenging and dynamic reverberant conditions of the dataset, with sources appearing, overlapping, and disappearing often in the testing set. Adding the dMOTa loss increases the $MOTa$ and $LR$ metrics further. Apart from improvements in $LE$ and $LR$, dMOTa improves trajectory consistency at the regressor outputs; something that is not captured by the $LE,LR$ metrics. Instead, this improvement is exemplified by the $IDS$ scores, which drop significantly when dMOTa is included.

For a comparative look with other systems on the same dataset, we include the top three systems of the DCASE2020 challenge, along with their reported challenge $LE,LR$ results in the development dataset. The proposed training strategy of multi-source regression SSL is competitive against those methods, with both $LE$ and $LR$ being on a similar range. Furthermore, the proposed DOAnet with differentiable tracking-based training is much simpler than these proposals in terms of complexity, and it achieves such results without relying on additional sound class information. However, it has to be noted that the comparison is qualitative, since the $LR$ and $LE$ scores in the challenge submissions are first computed between the target sound classes, and then averaged.

\section{Conclusions}
A method has been presented for end-to-end training of regression-based multi-source localizers that can handle realistic training data of time-varying varying source numbers, overlapping scenarios, and moving sources. Similarly, during inference and for the same dynamic acoustic conditions, the method achieves low localization errors, high localization detection  scores, and improved tracking performance between the multiple DOA regressors. The approach is competitive against state-of-the-art SELD systems, at a reduced complexity and without dependency on sound-type detection information. 

\bibliographystyle{IEEEtran}
\bibliography{refs}
\end{document}